# Training for Transport and Localization in Quantum System

Madhuri Mukhopadhyay[1]

[1]School of Mathematics and Natural Sciences, Chanakya University, Bangalore, India
(Dated: December 31, 2025)

Through periodic "Training" we can gradually buildup a reproducible responses in a disordered system where plasticity dominates over elasticity as is known in classical amorphous materials and soft matter [1–6]. Here we show that a similar concept can be extended to disordered *quantum* systems. Periodic electrical or mechanical driving of a disordered quantum-dot network can sculpt the effective Hamiltonian, producing either a low-energy transport valley that enhances exciton conduction, or a localized regime with many-body-memory-like behavior. Our results establish *training* as a new paradigm for creating functional order in disordered quantum matter.

Disordered quantum systems—such as random arrays of quantum dots, molecular complexes, or doped semiconductors—present a tantalizing tension: disorder tends to localize carriers or excitations, while coherent coupling and delocalization promote transport [7, 8]. In conventional paradigms, disorder is a fixed background, and one studies how to mitigate its deleterious effects via screening, phonons, or engineered coupling [9, 10]. But what if, instead, one could *train* the disordered quantum system—by repeated external driving or pulses—so that the disorder landscape itself evolves and becomes functional?

Training disordered matter is not a new idea in classical contexts. In mechanical and amorphous solids, researchers have shown that periodic deformation or "creep" cycles can sculpt the internal structure and memory of disordered networks, coupling a chosen "source" to a remote "target" response [1, 11, 12]. More broadly, memory formation in materials under cyclic driving has been explored in soft matter, glasses, and elastoplastic networks [5, 6]. The question we ask here is: *can analogous training (via electrical, mechanical, or field pulses) be applied to a quantum disordered system, so that its Hamiltonian self-organizes toward a more favorable form?*

Concretely, we consider a network of quantum dots or discrete sites with disordered on-site potentials $\epsilon_i$ and couplings $J_{ij}$. Through cyclic external driving (e.g., strain, gate pulses, or electric fields), the effective couplings or site energies may adjust plastically (or via microscopic rearrangements, charge redistribution, or local field reconfiguration) [13, 14]. Over many cycles, the system may "learn" a reproducible response path or energy channel. **(i) Trained transport valley / enhanced conduction.** Training may sculpt a low-energy valley in the coupling network, aligning $J_{ij}$ and lowering potential barriers, thereby enabling faster and more coherent transport along a reproducible channel. The signature of low energy channel and the coherent transport have been observed experimentally in some biological systems. [9, 16–19].

**(ii) Trained localization / emergent many-body**

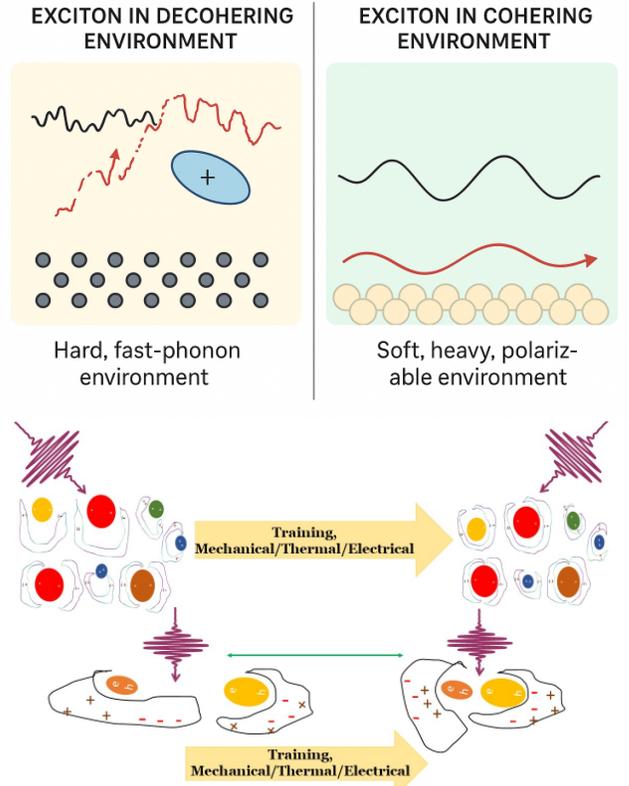

FIG. 1. **Upper Panel: Schematic of hard, fast Phonon environment, Lower Panel:Concept of training in a disordered quantum system.** Decoherence due to Phonon is higher, Soft, heavy Polarizable environment where decoherence due to Phonon is less. The Upper training shows localization is favored, where lower training shows better transport.The schematic parallels the training of amorphous solids [15], but here acts on a quantum Hamiltonian.

**memory.** Alternatively, if training induces local charging or increases effective Coulomb repulsion, certain regions may become more strongly localized, can display localization or many-body localization (MBL)-like signatures [20–22].

Many-body localization (MBL) itself is one of the most striking non-equilibrium phenomena in quantum mat-ter. Unlike thermalizing systems that obey the eigenstate thermalization hypothesis (ETH), MBL systems fail to equilibrate and retain memory of their initial states even at finite energy densities [23–25]. They exhibit logarithmic entanglement growth, suppressed transport, and characteristic level-statistics transitions from Wigner–Dyson to Poisson distributions [26]. However, most MBL theory treats *static* disordered Hamiltonians; the possibility of *learning or sculpting* the Hamiltonian under external cyclic driving remains largely unexplored.

We model the disordered quantum-dot ensemble as a Hubbard-type lattice whose parameters evolve under repeated electrical or mechanical training, a standard starting point for interacting localization studies [27, 28].

Strong disorder ($W \gtrsim t_0$) can drive many-body localization (MBL), producing long-lived local memory of initial states [29].

**Training and plasticity induced: Hamiltonian evolution.**— Nonlinear rectification under periodic driving generates a DC bias even for symmetric AC fields [30–32], enabling irreversible plastic rearrangements.

Pulse-induced activation follows a Kramers-type escape process [33, 34], with statistical multipliers depending on pulse correlations [30, 35].

**Quantum dynamics and observables.**— Exciton or electron transport diagnostics follow standard measures used to distinguish ergodic, diffusive, and localized regimes [36–38].

These diagnostics connect directly to experimental exciton transport and spectroscopy in molecular and quantum-dot systems [10, 16–19, 39].

We model the disordered quantum-dot (QD) ensemble as a Hubbard-type lattice whose parameters evolve under repeated electrical or mechanical training. The untrained Hamiltonian is

$$H_0 = -\sum_{\langle i,j \rangle} t_{ij}\, c_i^\dagger c_j + \sum_i \epsilon_i n_i + U \sum_i n_{i\uparrow} n_{i\downarrow}, \quad (1)$$

where $t_{ij} = t_0 e^{-\beta r_{ij}}$ denotes tunneling between dots separated by $r_{ij}$, $\epsilon_i$ are random on-site energies drawn from $[-W/2, W/2]$, and $U$ is the on-site Coulomb repulsion. Strong disorder ($W \gtrsim t_0$) drives many-body localization (MBL), producing long-lived local memory of initial states.

**Training and plastic evolution.**— When the QD lattice is embedded in a soft or visco-elastic host matrix, periodic electrical or mechanical pulses slightly modulate inter-dot distances, $r_{ij}(t) = r_{ij}^0 + \delta r_{ij}(t)$. A minimal visco-elastic relaxation law captures plastic adaptation,

$$\frac{dr_{ij}}{dt} = -\gamma \frac{\partial U_{\text{elastic}}}{\partial r_{ij}} + F_{\text{drive}}(t), \quad (2)$$

so that cyclic driving gradually reshapes the network. Consequently the electronic parameters evolve as $t_{ij} \to t_{ij}^{\text{trained}}$ and $\epsilon_i \to \epsilon_i^{\text{trained}}$. To emulate this process numerically, we apply simple update rules per cycle $n$,

$$t_{ij}^{(n+1)} = t_{ij}^{(n)} + \eta_J\, f(|\psi_i|^2, |\psi_j|^2),$$
$$\epsilon_i^{(n+1)} = \epsilon_i^{(n)} + \eta_\epsilon\, g(|\psi_i|^2), \quad (3)$$

where $f, g$ represent Hebbian-type activity couplings and $\eta_J, \eta_\epsilon$ are learning rates. This phenomenological "plasticity rule" efficiently generates trained Hamiltonians reproducing experimental trends of strain or electric-field cycling. Here we provides the theoretical framework connecting (i) nonlinear rectification under periodic driving, (ii) enhancement of activation (escape) rates under time-dependent work, and (iii) accumulation of irreversible plastic events that generate permanent structural changes.

These results justify the coarse-grained update rules and show how cyclic driving produces trained tight-binding parameters in a disordered microscopic landscape. We consider microscopic environments with local dipoles, polarizable bonds, or geometric orientations whose configuration determines model parameters such as site energies $\varepsilon_i$ and couplings $t_{ij}$. Let $A(t)$ denote a slow mesoscopic alignment variable; changes in $A$ map to

$$\varepsilon_i \to \varepsilon_i(A), \qquad t_{ij} \to t_{ij}(A).$$

Local plastic rearrangements modify $A$ in discrete increments but can be captured by a coarse-grained continuous variable $S(t)$ representing structural order.

We use the following characteristic times:

$$\tau_{\text{plastic}} \gg \tau_A \gg \tau_{\text{pol}} \gg \tau_{\text{drive}},$$

ensuring clean scale separation between rapid dipole response, slow mesoscopic alignment, and rare irreversible plastic events.

If the instantaneous equilibrium value of $A$ in field $E(t)$ is $A_\infty[E(t)]$, then a standard relaxation form applies:

$$\tau_A \dot{A}(t) = -\left(A(t) - A_\infty[E(t)]\right). \quad (4)$$

Expanding the nonlinear response,

$$A_\infty(E) = a_1 E + a_2 E^2 + a_3 E^3 + \cdots, \quad (5)$$

and taking a sinusoidal drive $E(t) = E_0 \sin \omega t$, the cycle average contains only even powers of $E$:

$$A_{\text{DC}} \equiv \overline{A(t)} \approx a_2 \frac{E_0^2}{2} + \mathcal{O}(E_0^4), \quad (6)$$

up to lag corrections depending on $\omega \tau_A$. Thus even a symmetric AC field generates a DC bias—a standard nonlinear rectification effect [30, 31].

**Chosen pulse families for simulation**

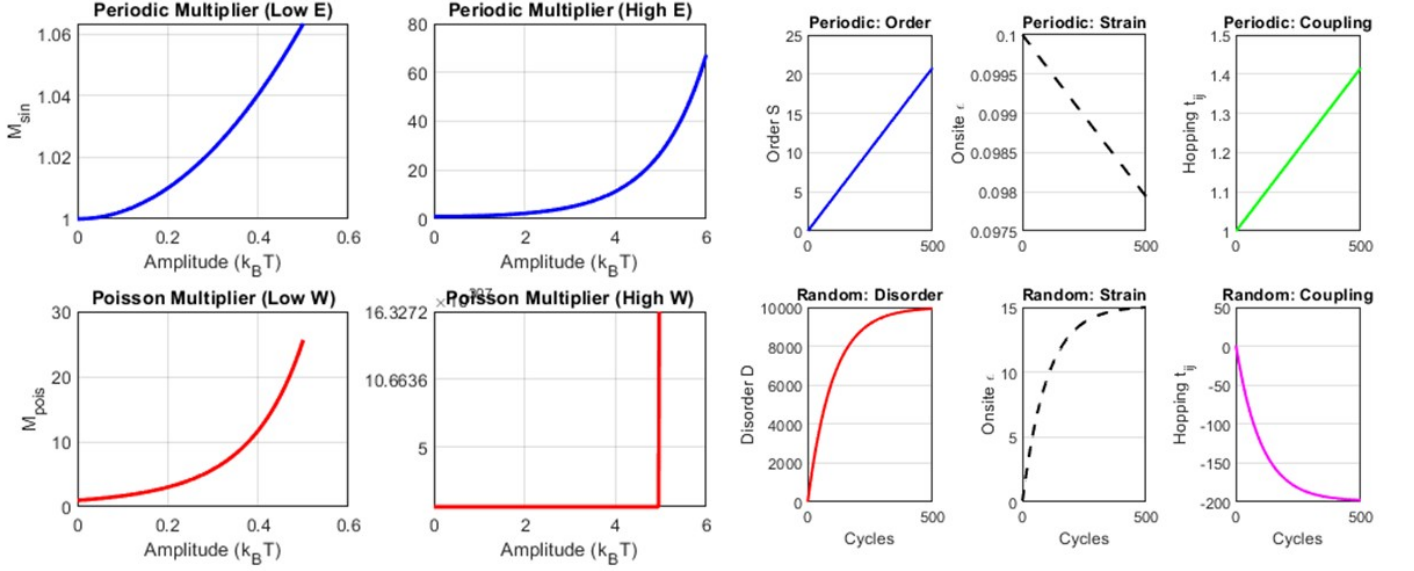

FIG. 2. **Training statistical control.** Periodic pulses generate smooth growth of order and uniform enhancement of $t_{ij}$, while Poisson pulses induce rapid disorder accumulation, strong onsite fluctuations, and suppression of hopping. The statistical multiplier thus can determine whether training creates a coherent transport channel or drives the system toward localization.

1. **Impulsive ($\delta$-like) pulses** to probe the limit of ultra-short, high-intensity kicks where activation is dominated by instantaneous barrier tilting.

2. **Random square pulses** to model finite-duration, sign-randomized pulses that efficiently activate slow degrees of freedom without global alignment.

3. **Poisson (random-timing) pulses** to represent a statistically clean model of noise-driven activation with analytically controllable event rates.

These three classes span the space of short, stochastic, and broadband drives, covering both impulsive and quasi-noise limits. All other pulse types lie between these three limiting behaviors and can be understood through interpolation.

**Pulse-induced activation: Kramers-type mechanism**

Let $x$ denote a local structural coordinate (bond rotation, hydrogen-bond configuration, segment orientation) with an intrinsic bistable potential $U(x)$ and a barrier height $\Delta U$ separating metastable states. An applied pulse performs instantaneous or time-dependent work $W(t)$ on this coordinate, modifying the effective barrier:

$$\Delta U(t) = \Delta U - W(t). \qquad (7)$$

The instantaneous escape rate follows the Kramers expression [33, 34]:

$$r(t) = r_0 \exp\left[-\frac{\Delta U - W(t)}{k_B T}\right]. \qquad (8)$$

For pulses satisfying $W(t) \neq 0$ only during short time windows, the cycle-averaged rate becomes

$$\bar{r} = \frac{1}{T}\int_0^T r(t)\,dt = r_0 e^{-\Delta U/k_B T}\left\langle e^{W(t)/(k_B T)}\right\rangle, \qquad (9)$$

where $\langle \cdot \rangle$ denotes the time or ensemble average over pulses.

For a $\delta$-pulse of amplitude $W_0$,

$$\left\langle e^{W(t)/(k_B T)}\right\rangle = 1 + \frac{W_0}{k_B T}\frac{\tau_\delta}{T} + O(W_0^2), \qquad (10)$$

where $\tau_\delta$ is the (short) effective duration.

For a random square pulse of amplitude $W_0$, duration $\tau_{\text{on}}$ and period $T$,

$$\left\langle e^{W(t)/(k_B T)}\right\rangle = \frac{\tau_{\text{on}}}{T}e^{W_0/(k_B T)} + \left(1 - \frac{\tau_{\text{on}}}{T}\right). \qquad (11)$$

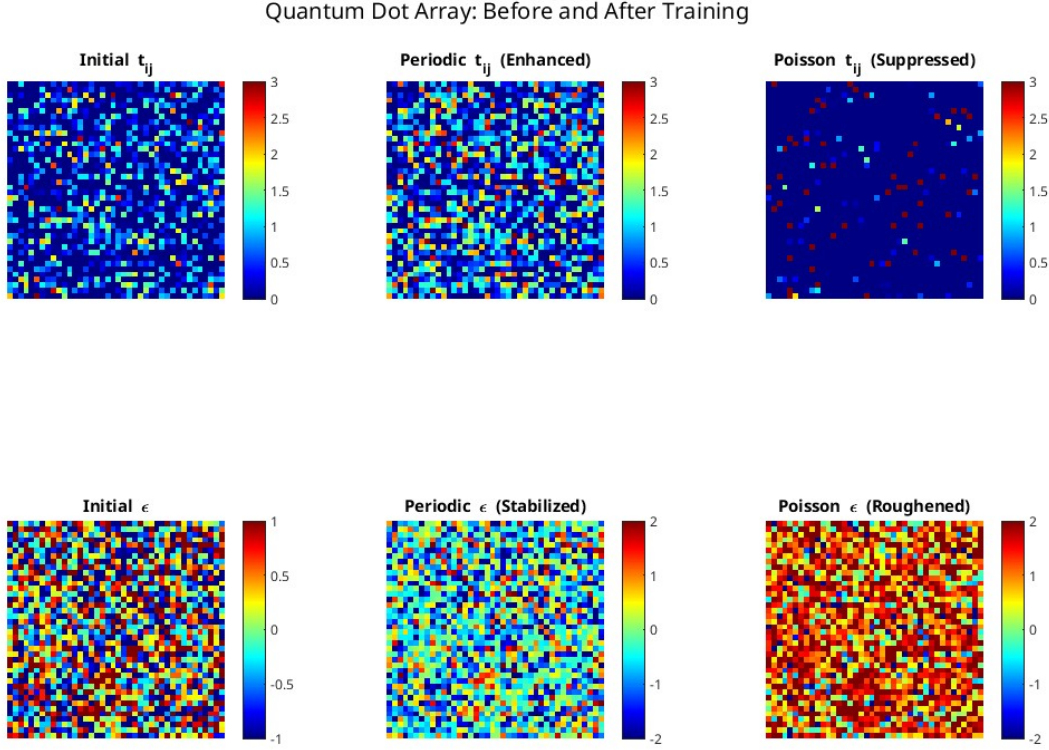

FIG. 3. **Evolution of coupling $t_{ij}$ and onsite disorder $\epsilon_i$ in a quantum-dot array before and after training.** Top row: spatial maps of hopping amplitudes. The initial untrained array displays random couplings; periodic driving selectively enhances $t_{ij}$ and forms a more connected, high-coupling transport network, whereas Poisson-type stochastic training suppresses most links, leaving only sparse connectivity indicative of localization. Bottom row: onsite potential landscape. Periodic training slightly smooths and stabilizes $\epsilon_i$, reducing large fluctuations, while Poisson-driven activation roughens the potential landscape and increases disorder amplitude. Together, the maps illustrate how the *statistics of training* alone determine whether the Hamiltonian becomes smoother and transport–supporting, or rougher and strongly localized.

For Poisson pulses with rate $\lambda$ and kick amplitude distribution $P(W)$, the well-known shot-noise result [30, 35] gives

$$\left\langle e^{W(t)/(k_B T)} \right\rangle = \exp\left[ \lambda \int (e^{W/(k_B T)} - 1) P(W)\, dW \right]. \quad (12)$$

In all cases, Eq. (9) shows that *the average escape rate grows with the statistical variance and amplitude of the pulse.* Hence appropriately chosen pulses efficiently trigger rare structural events.

**Disorder accumulation and evolution of structural variables**

Let $S$ denote a coarse-grained structural disorder parameter (e.g. variance of on-site potentials or orientation disorder). If each activated event changes $S$ by $\delta S$, then the mean-field evolution is

$$\dot{S}(t) = \delta S\, \bar{r}\, (1 - S/S_{\max}) - \gamma_S S, \quad (13)$$

where $(1 - S/S_{\max})$ imposes saturation and $\gamma_S$ describes any slow relaxation (often negligible in glassy media).

Substituting Eq. (9) yields an explicit dependence on the pulse statistics:

$$\dot{S}(t) = \delta S\, r_0 e^{-\Delta U/k_B T} \left\langle e^{W(t)/(k_B T)} \right\rangle (1 - S/S_{\max}) - \gamma_S S. \quad (14)$$

Equation (13) provides a unified framework for all pulse types: the choice of pulse merely changes the statistical factor $\langle e^{W/(k_B T)} \rangle$.

**Connection to trained disorder parameters**

The microscopic disorder modifies the tight-binding parameters via

$$\varepsilon_i^{\text{trained}} = \varepsilon_i^{(0)} - \alpha D_i(T_{\text{train}}), \qquad t_{ij}^{\text{trained}} = t_{ij}^{(0)} + \beta A_{ij}(T_{\text{train}}), \quad (15)$$

where $D_i$ and $A_{ij}$ depend on the final disorder state $S(T_{\text{train}})$. Hence disorder-enhancing pulses directly mod-ify transport and localization properties in the trained material.

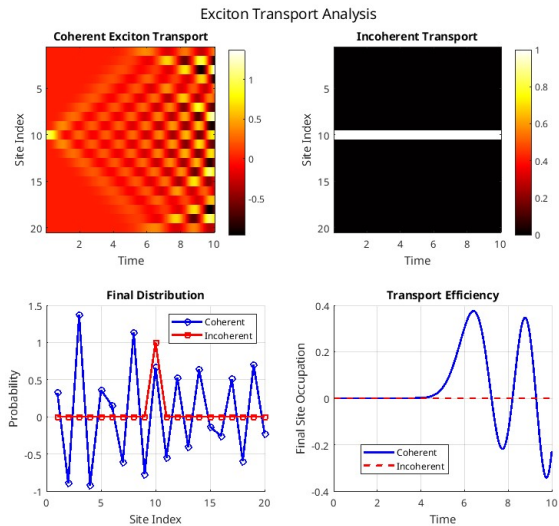

FIG. 4. **Exciton transport analysis: coherent vs. incoherent evolution.** The top panels show the time–site dynamics of exciton propagation for a trained system under (left) coherent transport and (right) incoherent (Poisson-type) dynamics. Periodic training produces a directed transport valley, enabling wave-like propagation across the lattice and clear interference structure. In contrast, stochastic Poisson training suppresses propagation and yields localization near the initial excitation site. Bottom-left: final site-occupancy distributions, highlighting ballistic-like spread for the coherent case and strong localization in the incoherent case. Bottom-right: transport efficiency, measured as occupation at the target boundary versus time, demonstrating oscillatory forward transfer under coherent evolution and negligible transport under incoherent dynamics. Together these results illustrate how periodic learning cycles sculpt coherent transport channels, while stochastic Poisson driving increases disorder and induces an MBL-like response.

The three selected pulse classes (impulsive, random square, and Poisson pulses) cover the full range of practical and theoretical possibilities for disorder-enhancing training. All pulses enhance structural activation through the universal Kramers-like rate (8), but differ in their statistical multiplier $\langle e^{W/(k_B T)} \rangle$, which controls the speed and nature of disorder accumulation.

**Quantum dynamics and observables.**— Exciton or electron dynamics follow the time-dependent Schrödinger equation

$$i\hbar \frac{d|\psi(t)\rangle}{dt} = H|\psi(t)\rangle, \qquad (16)$$

or, with dephasing,

$$\dot{\rho} = -\frac{i}{\hbar}[H,\rho] - \sum_i \gamma_i (n_i \rho + \rho n_i - 2 n_i \rho n_i), \qquad (17)$$

where $\gamma_i$ represents site-specific decoherence. Key diagnostics used below include the participation ratio IPR = $\sum_i |\psi_i|^4$ the mean-square displacement $\sigma^2(t) = \sum_i (r_i - r_0)^2 |\psi_i(t)|^2 \propto t^\alpha$, the level-spacing ratio

$$r_n = \min(\delta_n, \delta_{n+1})/\max(\delta_n, \delta_{n+1}) \qquad (18)$$

and the von Neumann entanglement entropy

$$S_A(t) = -\text{Tr}_A[\rho_A \ln \rho_A]. \qquad (19)$$

The variation of these quantities before and after training can quantify the crossover from diffusive to ballistic transport and from ergodic to MBL-like regimes, in the coming extended part those will be discussed in details.

In the following, we present numerical results for one- and two-dimensional lattices using Eqs. (1)–(17), demonstrating the emergence of trained transport valleys and localized memory pockets. Additional parameter scans, coupling maps, and scaling tests

In this brief version we have focused mainly on the conceptual framework and first-principles demonstration of how periodic training can sculpt transport-enhancing or localization-favoring Hamiltonians in disordered quantum systems. A full quantitative many-body localization (MBL) characterization—including level-statistics analysis ($r$-ratio), participation ratios, transport exponents, entanglement growth, and finite-size scaling across disorder strength and training cycles—will be presented in a forthcoming extended manuscript. That second part will include full numerical pipelines (ensemble generation → diagonalization → time evolution), statistical convergence analysis, and comparison with experimentally measurable spectroscopy signatures in trained quantum-dot films.

Here, we demonstrated that concepts originally developed in classical amorphous materials—namely plastic learning, cyclic preparation, and memory formation—admit a meaningful extension into quantum disordered systems. Under periodic electrical or mechanical pulses, the Hamiltonian of a disordered quantum-dot ensemble evolves along a learning trajectory, either opening low-energy transport valleys or enhancing effective Hubbard-type repulsion leading toward localized, memory-retaining phases. These results establish "training" as a controllable knob that can dynamically tune the balance between coherence and disorder—providing a promising route to programmable exciton transport, tunable localization, and emergent quantum functionality in soft-matter-embedded quantum materials.

### ACKNOWLEDGMENTS

We gratefully acknowledge insightful discussions from Dr. Himangsu Bhaumik, JNCASR, Bangalore, whose invited talk on the training behavior of glassy and amor-phous materials helped and inspired the analogy and conceptual direction of this work. We also thank the support, facilities and funding from the School of Mathematics and Natural Sciences, Chanakya University, and the colleagues for broader discussions and feedback during development of the theoretical model.